# Neutron Zeeman beam-splitting for the investigation of magnetic nanostructures

S.V. Kozhevnikov[1*], F. Ott[2,3], E. Semenova[4]

[1]Frank Laboratory of Neutron Physics, JINR, 141980 Dubna, Russian Federation
[2]CEA, IRAMIS, Laboratoire Léon Brillouin, F-91191 Gif sur Yvette, France
[3]CNRS, IRAMIS, Laboratoire Léon Brillouin, F-91191 Gif sur Yvette, France
[4]Condensed Matter Department, Faculty of Physics, Tver State University, 170002 Tver, Russian Federation

**Abstract**

The Zeeman spatial splitting of a neutron beam takes place during a neutron spin-flip in magnetically non-collinear systems at grazing incidence geometry. We apply the neutron beam-splitting method for the investigation of magnetically non-collinear clusters of submicron size in a thin film. The experimental results are compared with ones obtained by other methods.

## 1. Introduction

Achievements in nanotechnologies require new methods of nanostructures characterization. Neutron scattering is a powerful tool for the investigation of biological objects, polymers and magnetic systems. Polarized neutron reflectometry (PNR) is routinely used to probe magnetic structures in thin films (at scales in the range of 3-100 nm) [1]. Off-specular scattering [2] appears happens when the film contains in-plane microscopic structures in the direction along the beam propagation (for length-scales in the range 600 nm - 60 μm). Grazing Incidence Small-Angle Neutron Scattering (GISANS) can be used when the film has nanometric in-plane structures in the direction perpendicular to the beam path [3-6]. The momentum transfer takes place in this direction, the scattering plane is perpendicular to the incidence plane and the inhomogeneity scale is 3-100 nm.

Using together the methods of PNR, off-specular scattering and GISANS opens the way towards 3D-tomography of magnetic nanostructures. However, there are essential restrictions. These methods are model-dependent, average information over a surface and need periodicity or spatial coherence of inhomogeneities. Therefore, complementary application of direct methods may increase the data accuracy. Neutron methods for direct determination of the magnetic induction in thick films (thickness > 100 nm) are Larmor precession [7], neutron magnetic resonance [8] and the Zeeman spatial splitting of the neutron beam [9,10]. A review on these three methods can be found in [11].

In this communication, the beam-splitting method is applied for the direct determination of the magnetic induction inside non-collinear magnetic clusters in a Fe-Gd thin film. We present experimental data and compare it with other ones obtained by complementary methods such as PNR, Magneto-Optical Kerr Effect (MOKE), Vibrating Sample Magnetometry (VSM).

## 2. Zeeman spatial beam-splitting method

The phenomenon of Zeeman spatial beam-splitting at the boundary of two magnetically non-collinear media was predicted theoretically in [12] and observed experimentally in the geometry of reflection [13-15] and refraction [16-19]. The beam-splitting was registered in thin magnetically anisotropic films with domains [20-22], internal anisotropy in super-lattices [23,24] and clusters [25,26].

A detailed description of the beam-splitting method and data representation is provided in [10]. Here we simply recall the geometry of beam-splitting experiments. In Fig. 1a, the reflection and transmission through a thin magnetic film on a non-magnetic substrate is presented. The strong external field $B_0 \approx 1$ T in the air (the medium 0) is applied perpendicularly to the film surface. The vector of magnetization is directed parallel to the sample surface due to demagnetizing factor. Thus, the magnetic induction $B$ in the film (the medium 1) is directed under an angle $\chi$ to the vector of the applied magnetic field. The magnetic field in the nonmagnetic substrate (the medium 2) is equal to $B_0$ and the nuclear potential is $U$. The spin-flip occurs at the boundaries '0-1' and '1-2'. In this case, one observes three beams in reflection and three beams in refraction. Axis O$x$ is parallel to the film surface and axis O$z$ is perpendicular to the surface. The incident polarized beam with spin (+)/up or (−)/down enters under the incidence angle $\theta_i$. The final angle of reflected or refracted beam is $\theta_f$. The specular reflection takes place at $\theta_f = \theta_i$ but the spin-flip reflections (+−) and (−+) take place for $\theta_f \neq \theta_i$. The spin-flip probability depends on the angle $\chi$ between the vectors of magnetic induction $\mathbf{B}_0$ and $\mathbf{B}$ as $W \sim \sin^2 \chi$. For the theory see the references [12, 27]. The spin-flip probability as a function of the angle $\chi$ was measured experimentally in [16, 28]. The incident beam neutron wave vector is $k_i^2 = k_{iz}^2 + k_{ix}^2$. The wave-vector of the reflected beam is $k_f^2 = k_{fz}^2 + k_{fx}^2$ If the film magnetization is homogeneous along the boundary (along the $Ox$ axis) then the neutron velocity is unchanged along $Ox$, i.e. $k_{ix} = k_{fx}$.

---

[*] Corresponding author: kozhevn@nf.jinr.ru

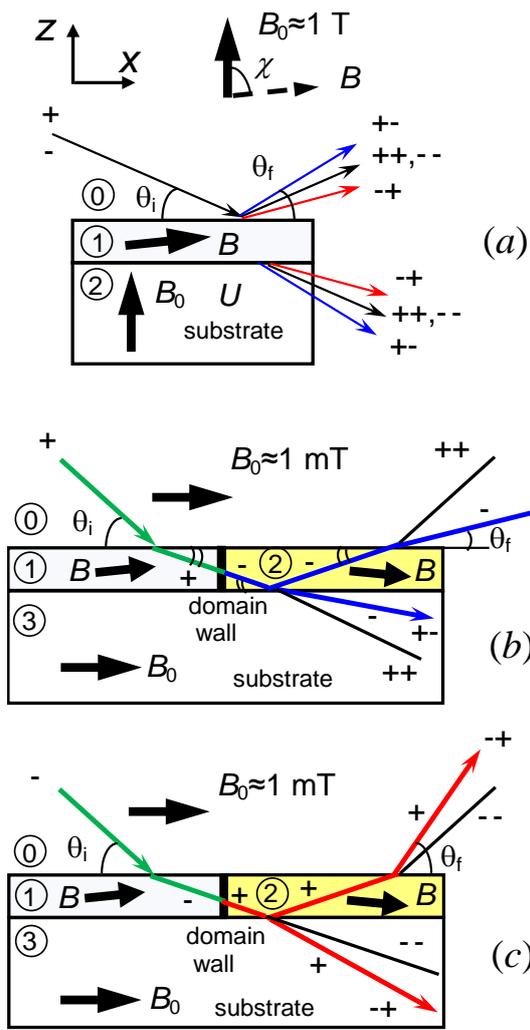

Fig. 1. Different beam-splitting geometries: (b) a thin magnetic film of induction $B$ on a nonmagnetic substrate of nuclear potential $U$ in a high perpendicular applied field $B_0 \approx 1$ T; (b) beam-splitting (+-) in a magnetic film with domain walls in a low parallel applied field $B_0 \approx 10$ mT; (c) the same for (-+) beam-splitting. The value of angular beam-splitting depends on the difference of the Zeeman energy during spin-flip at magnetically non-collinear boundary. The spin-flip probability depends on the angle $\chi$ between two vectors of magnetic induction in magnetically non-collinear media as $W \sim \sin^2 \chi$.

Let us denote $k_{iz} = k \sin \theta_i = p_i = \dfrac{2\pi}{\lambda}$ and $k_{fz} = k \sin \theta_f = p_f = \dfrac{2\pi}{\lambda}$, where $\lambda$ is the neutron wavelength. The kinetic energy for a neutron of wavelength 1.8 Å is equal to 25 meV. The change of the Zeeman energy in a field $B_0=1$ T is equal to $2\mu B_0=120$ neV (where µ is magnetic moment of neutron) which is negligible in comparison to the total kinetic energy but comparable with its $k_z$ component at small angles $\theta_i \approx 1°$. During a spin-flip, the change of Zeeman energy leads to a significant change in the $k_z$ component. This change in wave-vector corresponds to a measurable change in direction of the neutron beam.

The energy conservation law for the spin flip '+-' (where '+' is the spin of the incident beam parallel to the applied field $B_0$ and '-' is the spin of the transmitted beam antiparallel to the induction $B$ in the film) can be written as:

$$\frac{\hbar^2 p_i^2}{2m} + \mu B_0 = \frac{\hbar^2 (p_f^{+-})^2}{2m} - \mu B_0 + U \quad (1)$$

At transmission through the film there are three beams:

$$(\theta_f^{+-})^2 = \theta_i^2 + 2\mu B_0 \lambda^2 \frac{2m}{h^2} - U\lambda^2 \frac{2m}{h^2} \quad (2)$$

$$(\theta_f^{++})^2 = \theta_f^{--\,2} = \theta_i^2 - U \frac{2m}{h^2} \lambda^2 \quad (3)$$

$$(\theta_f^{-+})^2 = \theta_i^2 - 2\mu B_0 \lambda^2 \frac{2m}{h^2} - U\lambda^2 \frac{2m}{h^2} \quad (4)$$

For the reflected beam, it is $U=0$ in eqs. (1-4) and we have also three beams. One is the specular beam '++' and '--' and two others are off-specular beams:

$$(\theta_f^{+-})^2 = \theta_i^2 + 2\mu B_0 \lambda^2 \frac{2m}{h^2} \quad (5)$$

$$(\theta_f^{-+})^2 = \theta_i^2 - 2\mu B_0 \lambda^2 \frac{2m}{h^2} \quad (6)$$

Now we consider the particular case of the beam-splitting in a domain structure placed in a low magnetic field $B_0 \approx 1$ mT applied parallel to the surface of a thin film on a nonmagnetic substrate (Figs. 1b,c). Here the medium 0 is air, the medium 1 is a domain with the magnetic induction $B \approx 1$ T, the medium 2 is a domain with the same value of induction $B$ (for the sake of simplicity) but with a direction different from the domain 1. Between the non-collinear domains 1 and 2 there is a sharp domain wall where spin-flip takes place. The medium 3 is a nonmagnetic substrate. Let us consider the incident polarization '+' (Fig. 1b). The incident beam enters onto the surface under the incidence angle $\theta_i$. There are following conditions of the homogeneity along the boundaries: $k_{ix}^+ = k_{1x}^+$, $k_{1z}^+ = k_{2z}^-$ and $k_{2x}^- = k_{fx}^{+-}$. From the energy conservation law we can obtain the angles of the reflected off-specular beams:

$$(\theta_f^{+-})^2 = \theta_i^2 - 2\mu B \lambda^2 \frac{2m}{h^2} \qquad (7)$$

$$(\theta_f^{-+})^2 = \theta_i^2 + 2\mu B \lambda^2 \frac{2m}{h^2} \qquad (8)$$

The same for the transmission:

$$(\theta_f^{+-})^2 = \theta_i^2 - 2\mu B \lambda^2 \frac{2m}{h^2} - U\lambda^2 \frac{2m}{h^2} \qquad (9)$$

$$(\theta_f^{-+})^2 = \theta_i^2 + 2\mu B \lambda^2 \frac{2m}{h^2} - U\lambda^2 \frac{2m}{h^2} \qquad (10)$$

For the beams '++' and '--' the law of refraction is the same as eq. (3). One can see that the spin-flip '+-' in the domain structure corresponds to the spin-flip '-+' in the uniformly magnetized film and spin-flip '-+' in the domains corresponds to the spin-flip '+-' in the uniform film.

## 3. Experimental setup

The experiment was performed on the polarized neutron reflectometer SPN at the pulsed reactor IBR-2 (Frank Laboratory of Neutron Physics, Joint Institute for Nuclear Research, Dubna, Russia). Time-of-flight (TOF) method is used for the determination of neutron wavelength. The parts of the experimental setup are following: the curved long (5 m length) FeCo polarizer [29], non-adiabatic spin-flipper of Korneev type [30,31], the sample with a vertically oriented surface placed between the poles of the electromagnet, the second adiabatic radiofrequency spin-flipper[32] with a diameter of 100 mm, the multislit curved supermirror analyzer with working area 38×40 mm$^2$. The neutron beam is registered by a one dimensional $^3$He position-sensitive detector (PSD) with the working area 120(horizontally)×40(vertically) mm$^2$ and a spatial resolution of 1.5 mm [33]. The distance 'moderator-sample' was fixed as 29 m. The distance 'sample-detector' was 3 or 8 m. The width of the reactor pulse was 320 μs which corresponds to a neutron wavelength resolution of 0.02 Å for TOF base 37 m. The magnetic field was rotated with respect to the sample surface in the range 0-90°.

The polarization efficiencies of the polarizer and the analyzer were defined by 3P2S (3 polarizers and 2 spin-flippers) method [34]. The four reflectivities of an investigated sample $R^{++}, R^{--}, R^{+-}$ and $R^{-+}$ were extracted taking into account the calibration of polarizing efficiencies of the polarizer and the analyzer and the flipping efficiency of the spin-flippers. This procedure is described in work [35] for the new reflectometer REMUR which replaced the old reflectometer SPN in 2003. Some parameters of the reflectometer SPN can be found in the reference [35].

The sample was a bilayer Gd(5 nm)/Fe(100 nm)//glass(substrate) with the sizes 100(along beam)×60(width)×5(substrate thickness) mm$^3$. The initial motivation of the investigation of this sample was demonstration of neutron standing waves [36,37] using enhanced neutron absorption in the absorbing Gd layer. But during the experiment, the beam splitting in a low parallel external field was recorded on the position-sensitive detector. These experimental results are qualitatively discussed in [25,26]. In this communication we compare neutron data with other techniques to develop the neutron beam-splitting method for the investigations of magnetically non-collinear nanostructures.

## 4. Experimental results

### 4.1. MOKE data

To characterize magnetic properties of the thin magnetic film Gd(5 nm)/Fe(100 nm)//glass, we applied MOKE technique. It is the surface sensitive relative method. In Fig. 2, the hysteresis loop measured by MOKE in relative unites $M/M_s$ (where $M_s$ is the magnetization at saturation) is presented. The symbols correspond to the magnitudes of the applied magnetic field used in the neutron experiment. One can see that the film is a soft ferromagnetic with a low coercive field about 175 mT. The film is fast magnetized in the filed from 175 up to 300 mT. In the field below 175 mT the sample has a negative magnetization, and above 175 mT the film is positively magnetized. In the field about 175 mT the film is completely demagnetized. It is the macroscopic averaged information about magnetic properties of the sample surface. To extract more detailed information about microscopic properties of the magnetically non-collinear nanostructure in the film, we use polarized neutron reflectometry in grazing incidence geometry.

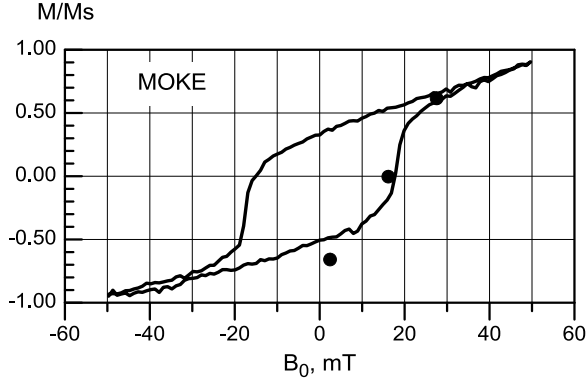

Fig. 2. Hysteresis loop measured by MOKE (line). The symbols mark the field value at which the neutron measurements are presented.

### 4.2. Perpendicular magnetic field

For the aim of spin-flip probability estimation, we performed the measurement in a field $B_0$=0.653±0.033 (T) applied under an angle of 80° with respect to the sample surface (see inset in Fig. 3). Here $\delta B_0$=0.033 T is the inhomogeneity of the applied field over the sample surface 60×100 mm$^2$ produced by the electromagnet poles. The magnetic film was magnetized in an in-plane magnetic field, then the applied field was turned to an angle of 80° with respect to the sample surface. Due to the shape anisotropy, the component of the magnetization perpendicular to the sample surface is compensated by the internal demagnetizing magnetic field $D\mathbf{M}$. The magnetic induction vector in the film is
$\mathbf{B} = \mathbf{B}_0 + (1-D)\cdot\mathbf{M} = \mathbf{B}_0 + \mathbf{M}_{//}$, where $\mathbf{M}_{//}$ is the component of the magnetization parallel to the sample plane.

The grazing angle of the incident beam was $\theta_i$=0.216° with an angular divergence of 0.005°. The distance 'sample-detector' was 8 m. In Fig. 3, the two-dimensional intensity map $I(\lambda, \theta_f)$ is presented for the mode up-down (Fig. 3a) and down-up (Fig. 3b). These raw data include a contaminating part of the neutrons with opposite spin due to the imperfection of polarization efficiency of the polarizer. The indices (+-) in Fig. 3a and (-+) in Fig. 3b denote the beams which experienced a Zeeman splitting. The indices correspond to the spin direction with respect to the direction of the applied field and were defined by total polarization analysis. One can see that this beam-splitting corresponds to the geometry in Fig. 1a. The key directions in Fig. 3a,b are the following: horizontal lines: specular reflection at $\theta_f$=0.216°, the horizon at $\theta_f$=0° and the direct beam at $\theta_f$=-0.216°. The (+-) reflected beam is above the specularly reflected beam (Fig. 3a). The (-+) reflected beam is in the region between the specularly reflected beam and the horizon (Fig. 3b) and reaches the horizon at a critical wavelength about 4.0 Å. For these neutrons, the $k_{iz}$ component is not sufficient to overcome the Zeeman energy change during a spin-flip. All refracted beams are scattered in the region between the direct beam direction and the horizon.

The suitable coordinates for the parameters determination are $(p_i^2 - p_f^2, p_i^2 + p_f^2)$ presented in Fig. 3c (up-down mode) and Fig. 3d (down-up). The diagonal dashed line is the horizon. The vertical line crossing the zero of abscises coordinate is the specularly reflected beam. The left vertical line (Fig. 3c) at -4.0·10$^{-3}$ nm$^{-2}$ is the reflected beam (+-) and the right vertical line at 3.8·10$^{-3}$ nm$^{-2}$ is the reflected beam (-+). The difference between these two values corresponds to the Zeeman energy $4\mu B_0$. The value of the applied field extracted from the Zeeman beam-splitting at reflection is $B_0$=0.672±0.052 (T). Here, the accuracy $\delta B_0/B_0$=7.8 % is estimated from the angular resolution $\delta\theta/\theta$=5.3 % and the wavelength resolution $\delta\lambda/\lambda$=1.0 %. The horizontal line crossing the zero of the ordinates axis is the direct beam line. Other horizontal beams are (+-) at 0.95·10$^{-3}$ nm$^{-2}$ in Fig. 3c and (-+) at 8.95·10$^{-3}$ nm$^{-2}$ in Fig. 3d. From the difference of these values, the parameter $B_0$=0.706±0.055 (T) is defined. These representations use the relative positions of the spin-flip beams and its difference give the magnetic parameter. Thus, an angular off-set which would simply shift the whole picture or an offset in the nuclear potential value do not contribute to the magnitude of $B_0$. The nuclear potential defined from the down-down intensity at refraction in instrumental coordinates (Fig. 3a) is $U$=96±8 (neV). And the same parameter defined from the squared normalize coordinate (Fig. 3c) is $U$=103±8 (neV). These values are equal to each other within error bar. In Fig. 3e, the specular reflectivities $R^{++}$ (closed symbols) and $R^{--}$ (open symbols) are presented as a function of neutron wavelength.

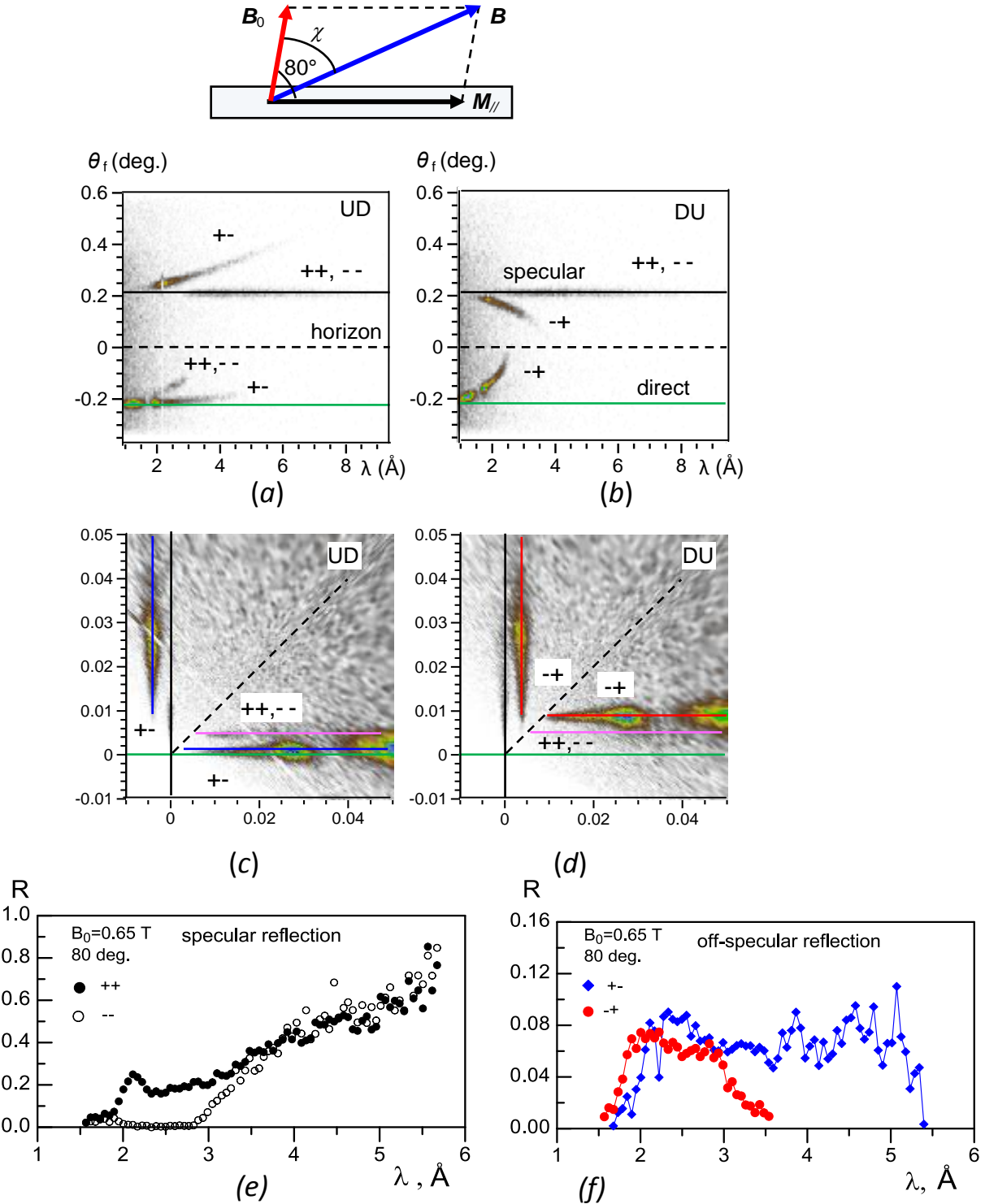

Fig. 3. The results in the high perpendicular applied field $B_0$=0.653 mT. (inset) Geometry of a magnetically non-collinear system. The two-dimensional map of intensity in the instrumental coordinates $(\theta_f, \lambda)$ for the mode up-down (a) and down-up (b). The two-dimensional map of intensity in the normalized squared coordinates for the mode up-down (c) and down-up (d). (e) Specular reflectivity '++' (closed symbols) and '--' (open symbols). (f) Off-specular reflectivity '+-' (rhombi) and '-+' (circles).

On the (++) reflectivity one can see a minimum at 2.5 Å which corresponds to the absorption in the Gd layer enhanced by a standing wave. Also, absorption takes place at long wavelengths and reflectivities does not reach a plateau $R=1$ of total reflection.

The spin-flip reflectivities $R^{+-}$ (rhombi) and $R^{-+}$ (circles) are presented in Fig. 3f. One can see the critical wavelength about 4.0 Å for (-+) reflectivity. For both spin-flip modes, the maximum of spin-flip reflectivity is 0.08. The spin-flip coefficients at the beam-splitting in the perpendicular field were obtainedexperimentally for reflection in [38,39] and for transmission in [28].

### 4.3. Parallel magnetic field

In a second configuration a low magnetic field $B_0$ was applied parallel to the sample surface. The detector was placed at a distance of 3 m from the sample and the incident collimating slit were open to $\delta\theta/\theta=20\%$. The neutron flux was increased by a factor 3.

The geometry of experiment is presented in Fig. 1b,c. We suppose that in the film there are clusters with a vector of magnetic induction $\boldsymbol{B}_1$ non-collinear with the magnetic induction $\boldsymbol{B}$ in the film.

In Fig. 4, the neutron data in a low parallel magnetic field are presented. The upper, middle and lower rows of panels correspond to the applied parallel field 1.8, 16.3 and 27.7mT, respectively. In Figs. 4a-f, the two-dimension maps of the neutron intensity $I(\lambda,\theta_f)$ are shown. The left and right columns correspond to up-down and down-up modes, respectively. The angle of the incident beam was 0.242°. The specular non spin-flip reflectivities $R^{++}$ (bold line) and $R^{--}$ (thin line) are shown in Figs.4 g, i, k. The spin-flip reflectivities $R^{+-}$ (rhombi) and $R^{-+}$ (circles) are presented in Figs. 4h,l (off-specular) and 4j (specular).

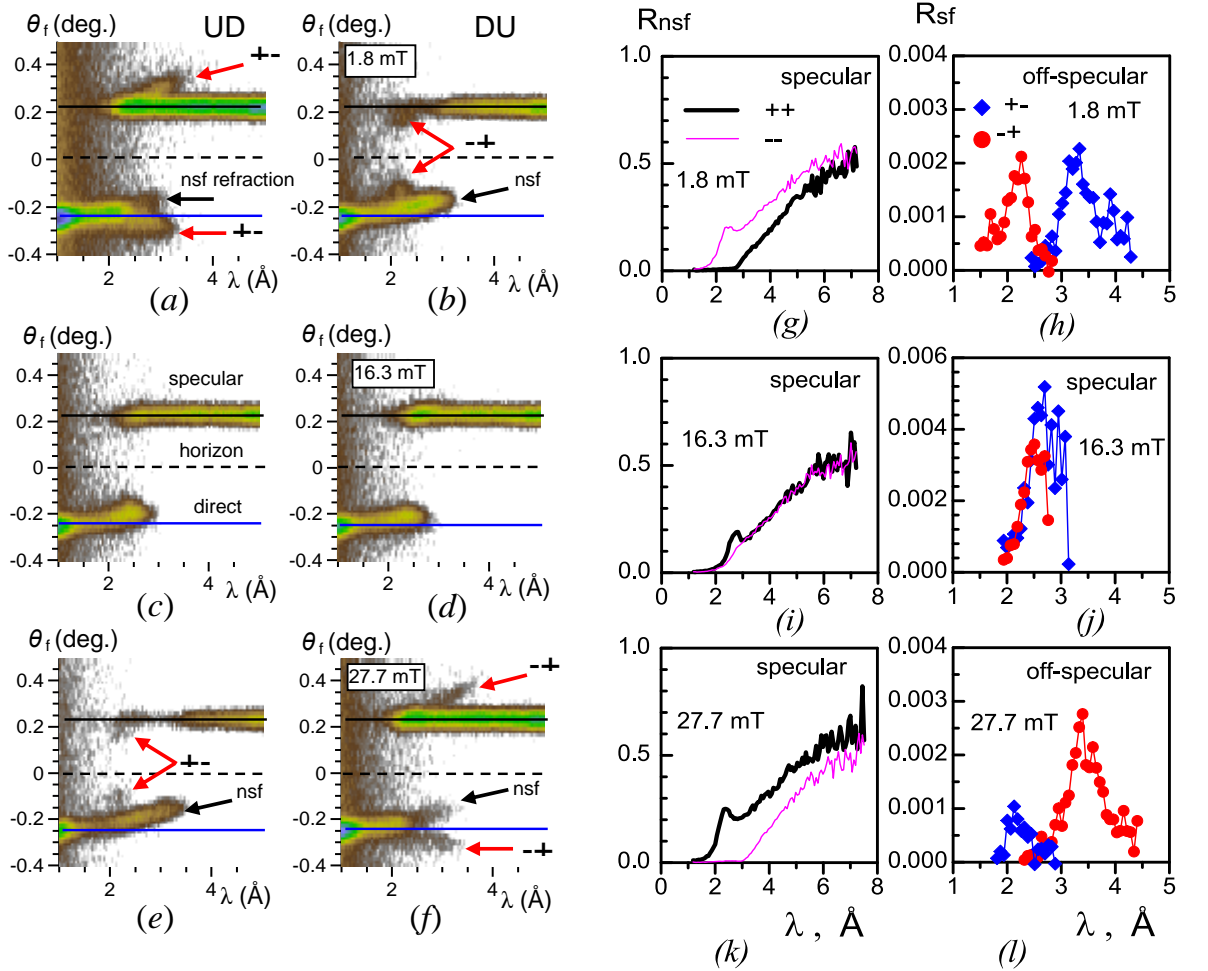

Fig. 4. The neutron data in a field parallel to the film surface. The upper row of panels corresponds to the field 1.8 mT and negative magnetization, the middle row to the field 16.3 mT and demagnetized state and the bottom row corresponds to the field 27.7 mT and positive magnetization. The left and right columns of two dimensional intensity maps (a-f) correspond to the modes up-down and down-up, respectively. The left and right columns of reflectivities (g-l) correspond to specular non spin-flip and spin-flip (off-specular for 1.8 and 27.7 mT and specular for 16.3 mT) reflectivities, respectively.

The sample was saturated in a high negative field and then the positive increase fields were applied (following the 4th quadrant of the hysteresis loop).

In the field of 1.8 mT beam-splitting is observed in up-down (Fig. 4 a) and down-up (Fig. 4b) modes. The arrows with the indices (+-) and (-+) indicate the corresponding beams for reflection and refraction. The arrow 'nsf' indicates the refracted beam 'non spin-flip'. One can see that off-specular scattering corresponds to the geometry of domain wall in Figs. 1b,c if we replace *B* by (-*B*). This fact is confirmed by the reflectivities in Fig. 4g which indicate the negative magnetization of the film because of the critical neutron wavelength of total reflection is less for (--) than for (++): $\lambda_c^{--} < \lambda_c^{++}$. For a positive magnetization it must be opposite: $\lambda_c^{--} > \lambda_c^{++}$. The off-specular spin-flip reflectivities maxima 0.002 consists of about 2.5 % of the off-specular spin-flip reflectivities in the high perpendicular field (Fig. 3f).

In the field of 16.3 mT, there is no beam-splitting (Figs. 4c,d). The sample is demagnetized and the non spin-flip specular reflectivities are merged (Fig. 4i).

The spin-flip specular reflectivities maxima are about 0.004 that consists of about 5.0 % of the off-specular spin-flip reflectivities in the high perpendicular field measurement.

In the field of 27.7 mT (Figs. 4e,f), there is a beam-splitting corresponding to the spin-flip at domain walls in Figs. 1b,c. The specular non spin-flip reflectivities (Fig. 4k) indicate a positive magnetization of the sample. In Fig. 4l, the off-specular reflectivity (+-) has a maximum of about 0.0025 and the reflectivity (-+) has a maximum of about 0.001.

Two-dimensional intensity map of (-+) mode in the instrumental coordinates $I(\lambda, \theta_f)$ at different values of the applied parallel field $B_0$ are shown in Fig. 5. The upper (red) box above the specularly reflected beam marks the integration region of reflected beam in the off-specular region and the lower (black) box below the direct beam indicates the integration region for spin-flip beam at refraction. The system is the most magnetically disordered at 18.1 mT. At 437 mT, the system is totally magnetically collinear, the spin-flip and the beam-splitting are absent.

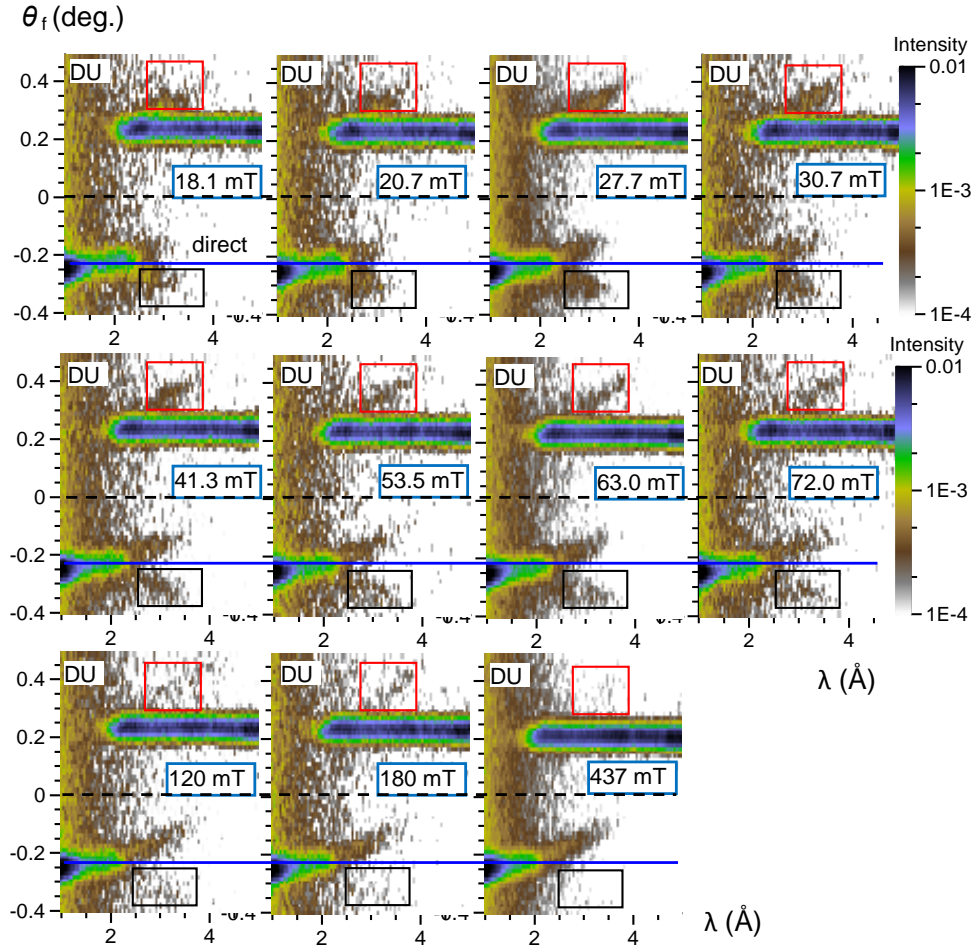

Fig. 5. Two-dimensional map of neutron spin-flip intensity of down-up mode in instrumental coordinates for different applied magnetic fields. Upper (red) and bottom (black) boxes mark the integration regions for reflection and refraction, respectively.

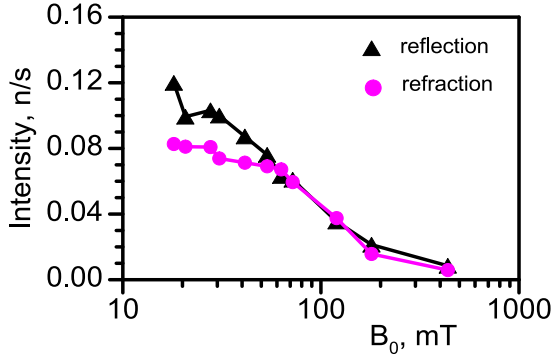

Fig. 6. The integrated spin-flip intensity at beam-splitting for reflection (triangles) and refraction (circles) as a function of the applied magnetic field value in the logarithm scale.

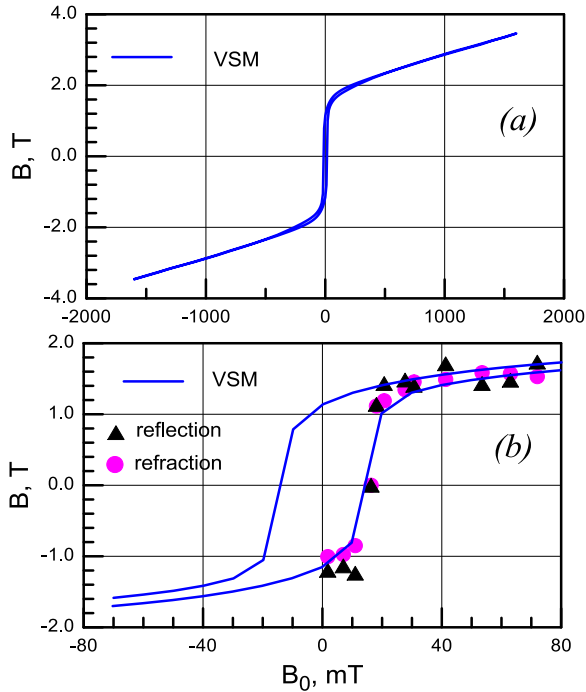

Fig. 7. Hysteresis loop obtained using: (a) VSM (line) in the full interval of the applied field; (b) VSM (line) and beam-splitting (triangles and circles correspond to reflection and refraction, respectively) in the short interval of the applied field.

In Fig. 6, the integrated intensity of the split beams in the mode up-down (in boxes in Fig. 5) is shown as a function of the applied magnetic field (in the logarithmic scale). The triangles and the circles correspond to the reflected and refracted beams, respectively. One can see that the spin-flip intensity decreases when increasing the applied field from 18.1 to 437.0 mT where the system is magnetically collinear.

In Fig. 7a the magnetic induction $B$ in the sample obtained by VSM is shown as a function of the applied magnetic field $B_0$ in the broad interval. The magnetic induction $B$ in the sample (derived from the beam-splitting value in Fig. 5) is presented in Fig. 7b as a function of the applied magnetic field $B_0$ in the narrow interval. The triangles and the circles correspond to the induction derived from the reflection. For the refraction, the value of the nuclear potential of the float glass substrate $U=96$ neV was used. The solid line is the hysteresis loop obtained by VSM. One can see that the neutron data are close to the VSM results. We can conclude that in two adjacent non-collinear media the magnetic induction value is equal to the magnetic induction in the Fe layer.

## 5. Microscopy data

To observe in-plane magnetic structures of the film, the Bitter technique was used. A colloid of small magnetic particles was placed on the film surface. The colloid is concentrated at areas with a maximal gradient of magnetic fields and makes these interfaces visible using optical microscopy.

The images of the magnetic clusters in different applied fields are presented in Fig. 8. The arrow indicates the direction of the applied in-plane magnetic field. One can see that the images in the demagnetized state without applied field $B_0=0$ and in the applied field $B_0=16.3$ mT are similar to each other. At $B_0=0$ in the black box of $10\times10$ μm$^2$ one can see spots of a diameter in the interval 0.3-0.7 μm (the single line inside the box has the length 2 μm). In this box, we can see 28 spots. Taking the average diameter 0.5 μm, we can estimate the part of these regions from the marked box area as $28\cdot3.14\cdot0.25^2/100=0.055$ or 5.5 %. This value coincides with the estimation from comparison of spin-flip reflectivities measured in a 16.3 mT parallel and a 0.653 T perpendicular fields (Figs. 4j and 3f, respectively).

In the field 18.1 mT, the image is close to the demagnetizing state. Starting from the field 27.7 mT, the picture is changed. The spots lengthen in the direction along the applied field and almost disappear in high magnetic fields (63.0, 120 and 150 mT). This effect can be explained by the fact that on the clusters, the volume of magnetic colloids is big. Therefore the force of the applied magnetic field pulling the magnetic particles in the colloid is also big. In this case the magnetic colloid is dispersed along the applied magnetic field. The domain wall thickness is not big and the mass of the magnetic powder particles in the colloid is not big enough to be dispersed by the applied magnetic field.

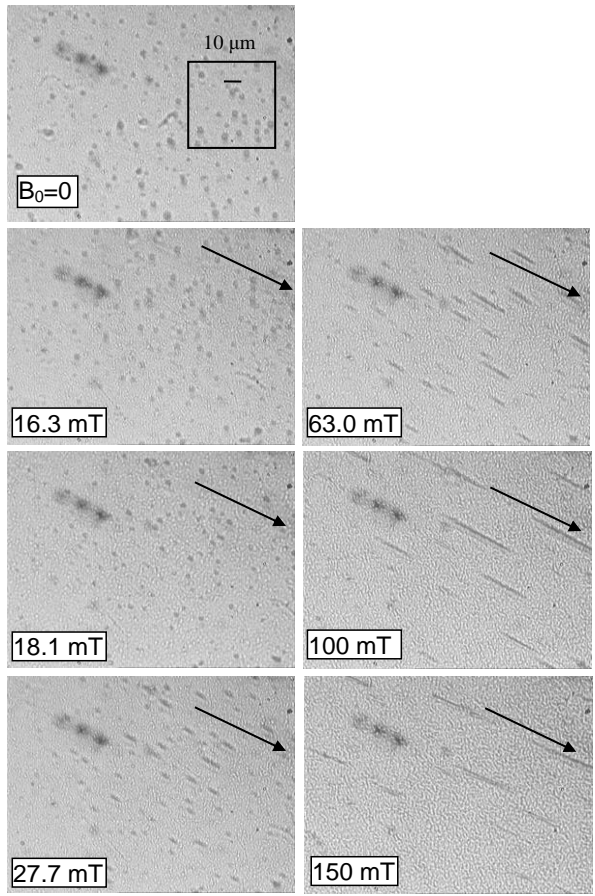

Fig. 8. The Bitter technique image for different values of magnetic field applied along the arrow.

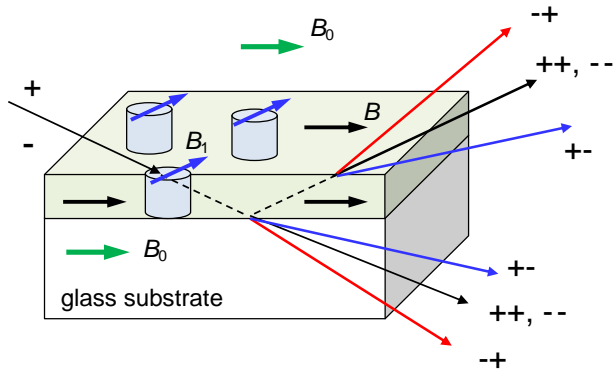

Fig. 9. The proposed model of the magnetically non-collinear clusters in the FeGd film on the float glass substrate. The neutron spin-flip occurs on the perpendicular boundary 'cluster-film' and the beam-splitting takes place at the transmission of the neutrons from the upper boundary of the film (reflected beam) and from the bottom boundary of the film (refracted beam).

## 6. Discussion

From the obtained results we can discuss the model of the structure in the Fe-Gd film. We propose that magnetic clusters correlated with the structure of the float glass substrate are formed in the Fe film. In Fig. 9 the model is presented. The structure of the float glass substrate produces magnetic clusters in the Fe-Gd film. The magnetic induction in the clusters is non-collinear with respect to the induction in the film in low applied magnetic fields. When the applied magnetic field becomes high, the induction in the clusters turns and becomes parallel to the induction in the entire film.

In [40] the defects in float glass substrates were investigated and 'craters' of diameter 0.6 µm were observed. It is very close to the situation in our sample. Thus we can suppose that the structural imperfections in the float glass substrate lead to the magnetic clusters in the deposited magnetic film with the same size.

In the investigated sample the enhanced absorption of neutrons is observed. The reason is Gd absorbing layer and the neutron standing wave formation. It is difficult to apply the conventional Polarized Neutron Reflectometry to this system. The reflectivity curves are deformed by the neutron absorption. The total reflection plateau is not 1 and the critical edge is also shifted. It is thus difficult to fit correctly the experimental data.

Magnetic clusters make up 5 % of the sample surface area. It is impossible to extract such low effect from the specular reflectivity. In such case, the beam-splitting method is the more sensitive method. In off-specular region we register only spin-flipped neutrons. Consequently, the level of background is very low. Therefore it is possible extract very small effects.

The beam-splitting method can be potentially applied for the investigations of magnetically non-collinear structures as for example patterned magnetic films or domains. In this case beam-splitting may be considered as a method of microscopic magnetometry. This complementary direct information may be very useful for the investigation of magnetic nanostructures.

## 7. Conclusions

We have investigated the magnetically non-collinear structure in a bilayer Fe-Gd film on a float glass substrate. The spatial neutron beam-splitting was observed in grazing incidence geometry at the low magnetic fields applied parallel to the film surface. Using Bitter technique, small magnetic clusters with a diameter of about 0.5 μm were observed in the film. The magnetic induction inside the magnetically non-collinear clusters was extracted by the beam-splitting method. This magnetization curve corresponds to the hysteresis loop measured by macroscopic magnetometric methods MOKE and VSM for the entire film. The neutron spin-flip probability measurements show that the vector of magnetic induction in non-collinear clusters becomes collinear to the magnetic induction in the film for applied field of about 400 mT. The beam-splitting method can be applied for the investigation of magnetically non-collinear nanostructures like as patterned films, domains for perpendicular magnetic recording, etc.

**Acknowledgements**

This work has been supported by the French project IMAMINE 2010-09 T. The authors are thankful to V.L. Aksenov, Yu.V. Nikitenko, J. Major, and F. Radu for fruitful discussions and help. The authors acknowledge A. Thiaville (Université Paris-sud, Orsay, France) and the researchers from Chair of Magnetism (Physical Faculty of Moscow State University, Russia) for the sample characterization by Kerr Effect.